\documentclass[11pt]{article}
\usepackage{moriond,epsfig}

\def\Journal#1#2#3#4{{#1} {\bf #2}, #3 (#4)}

\def\NPB{{\em Nucl. Phys.} B}
\def\PLB{{\em Phys. Lett.}  B}
\def\PRL{\em Phys. Rev. Lett.}

\def\ra{\rightarrow}

\def\be{\begin{equation}}
\def\ee{\end{equation}}
\def\bea{\begin{eqnarray}}
\def\eea{\end{eqnarray}}

\newcommand{\pt}    {\mbox{p$\rm _T$}}
\newcommand{\met}   {\mbox{$\rm E_T\hspace{-0.5cm}/\hspace{0.3cm}$}}

\newcommand{\fb}    {\mbox{$\rm \, fb\hspace{-0.08cm}^{-1}$}}
\newcommand{\GeVcc} {\mbox{$\rm \, GeV$}}

\newcommand{\GeV}   {\mbox{$\rm \, GeV$}}
\newcommand{\TeV}   {\mbox{$\rm \, TeV$}}
\newcommand{\Rinv}  {\mbox{$\rm R\hspace{-0.05cm}^{-1}$}}
\newcommand{\LR}    {\mbox{$ \rm\Lambda R$}}
\begin{document}
\vspace*{4cm}
\title{SEARCHES FOR NEW PHYSICS WITH~LEPTONS~IN~THE~FINAL~STATE~AT~THE~LHC}

\author{ M. KAZANA }

\address{
(on behalf of the ATLAS and CMS Collaborations) \\
 Laborat\'orio de Instrumenta\c{c}\~ao e F\'isica Experimental de Part\'iculas\\
 Av. Elias Garcia 14, 1$^{\circ}$, 1000-149 Lisboa, Portugal}

\maketitle\abstracts{
Final states including leptons are most promising to detect early signs of
new physics processes when the Large Hadron Collider will start
proton-proton collisions at the centre of mass energy of 14\TeV. The reach
for Supersymmetry and Extra Dimension models for integrated luminosities
ranging from 1 to 10\fb~is reported. Preliminary results indicate that
already with 1\fb~of data new phenomena can be detected. 
}

By the end of 2008 the ATLAS\,\cite{atlas} and CMS\,\cite{cms} experiments
at the LHC expect to collect between 0.5 and 1\fb~of data each, which
should make possible the first searches for new phenomena over the
Standard Model (SM) background. All such searches would require the
precision measurement of the SM processes with detailed understanding of
the detector performance, reconstruction algorithms and triggering.
Leptons, electrons and muons, have better reconstruction efficiency and
energy resolution than taus, jets and missing transverse energy (\met).  
They also provide a clean triggering and a high background reduction.
Above all, leptons may indicate signatures of new physics, such as decays
of massive strongly interacting particles to leptons accompanied by jets
and \met~and decays of new massive resonances to di-lepton pairs.  In this
report, the preliminary discovery limits for Supersymmetry (SUSY) and
Extra Dimension (ED) models estimated over a wide range of parameter space
are presented.

\vspace{-.2cm}
\section*{Inclusive SUSY reach with leptons}

Supersymmetry\,\cite{Martin:1997ns} is described by models which provide a
realistic SUSY-breaking scheme. One of the general approaches is given by
the Minimal Supergravity\,\cite{Alvarez-Gaume:1983gj} (mSUGRA) model with
only 5 free parameters:  a common scalar {\small $(m_0)$} and fermion
{\small $(m_{1/2})$} masses , a trilinear coupling {\small $(A_0)$} and
Higgs sector parameters {\small $(tan\beta,sgn\mu)$} at the Grand
Unification (GUT) scale. In mSUGRA, assuming
R-parity\,\cite{Martin:1997ns}, new supersymmetric particles are produced
in pairs and the lightest one (LSP) is stable and neutral. At the LHC, the
SUSY production is dominated by strongly interacting squarks and gluinos
($M_{SUSY}$\,$\sim$\,$m_{\widetilde{q},\widetilde{g}}$), which have long
decay cascades with the jet emission. The cascade ends with the LSP, which
is not detected. Therefore, a generic supersymmetry signature is a
multi-jet final state with large \met. The main backgrounds are $QCD$ and
$t\bar{t},~W,~Z$ with $QCD$-jet associated production processes, which
should be estimated by using an exact LO evaluation of partonic matrix
elements matched with parton showers at the hard process
scale\,\cite{Mangano:2002ea}.  In SUSY cascades, leptons are produced in
decays of charginos or neutralinos (e.g.{\small 
$\widetilde{\chi}^0_2\rightarrow \widetilde{\chi}^0_1 l^{+}l^{-}$,
      $\widetilde{\chi}^{+}_1\rightarrow \widetilde{\nu}_l l^{+}$}) and
the final state consists of $n$$\leq$4 leptons (+jets+\met). Pairs of
leptons can have the same or opposite sign (SS or OS).  Considering
signatures with at least one lepton, substantially reduces the $QCD$
background. The inclusive SUSY searches employ the following strategy.
First, experimental signatures are studied for a limited number of test
points of mSUGRA parameter space (in {\small $m_0$-$m_{1/2}$} plane for
fixed {\small $A_0$, $tan\beta$, $sgn\mu$}) using the full detector
simulation and reconstruction (S\&R) software. Next, the results are
extended to other points of the parameter space using fast S\&R.  In order
to obtain the best signal to background ($S/B$) ratio the SUSY selection
cuts are optimised for each point. The expected discovery reach is
evaluated for parameter sets having at least five standard deviation
(5$\sigma$) signal significance\,\cite{PTDR2}.

 \begin{figure}[t]
\begin{tabular}{rr}
\vspace{-.1cm}
 ATLAS preliminary ~~~~~~~~~~~~~~~~~~~& \\
\vspace{-5cm}
\epsfxsize=7.2cm \epsfysize=4.9cm  \epsfbox{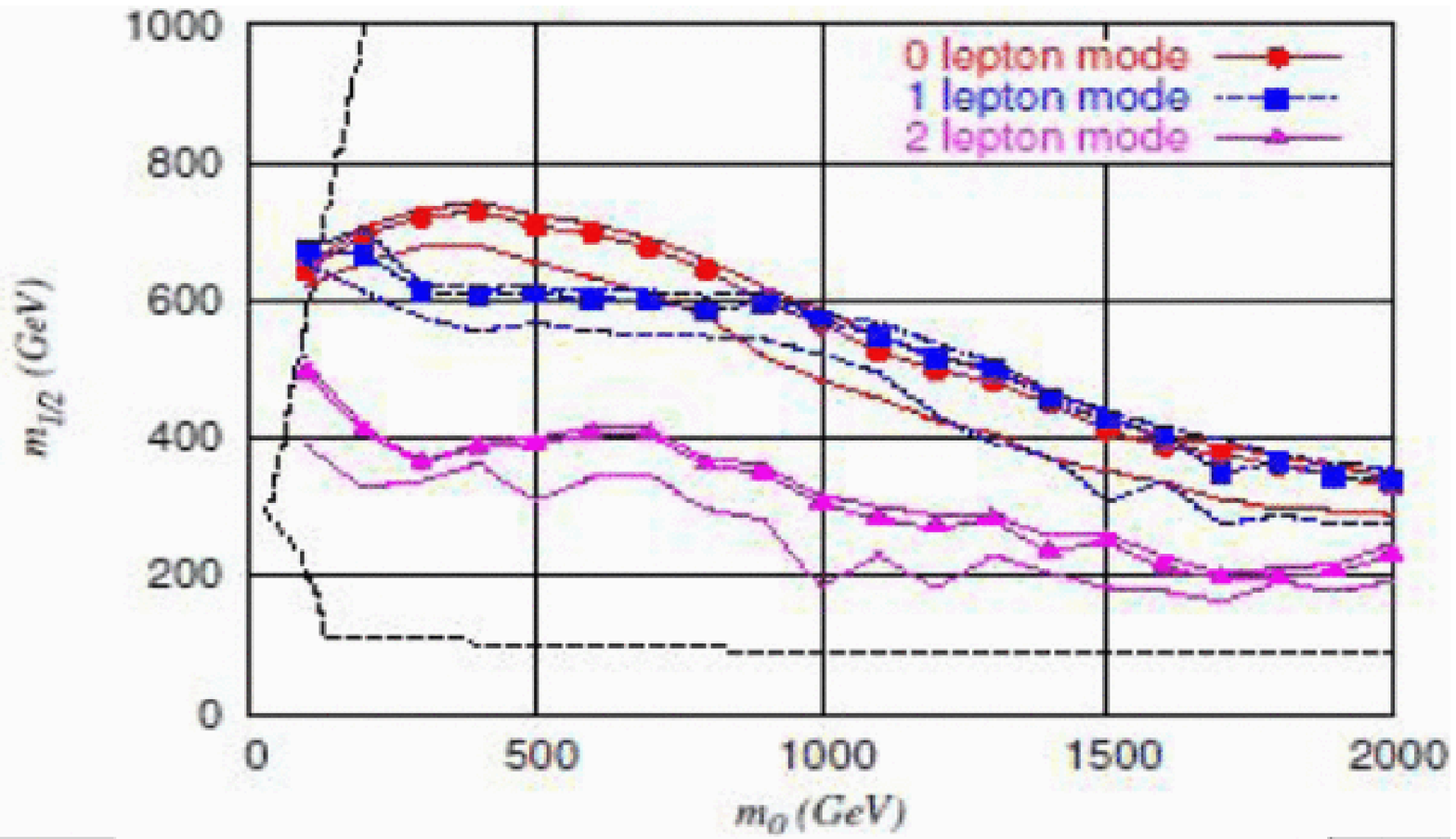}~~ &
\epsfxsize=7.2cm  \epsfbox{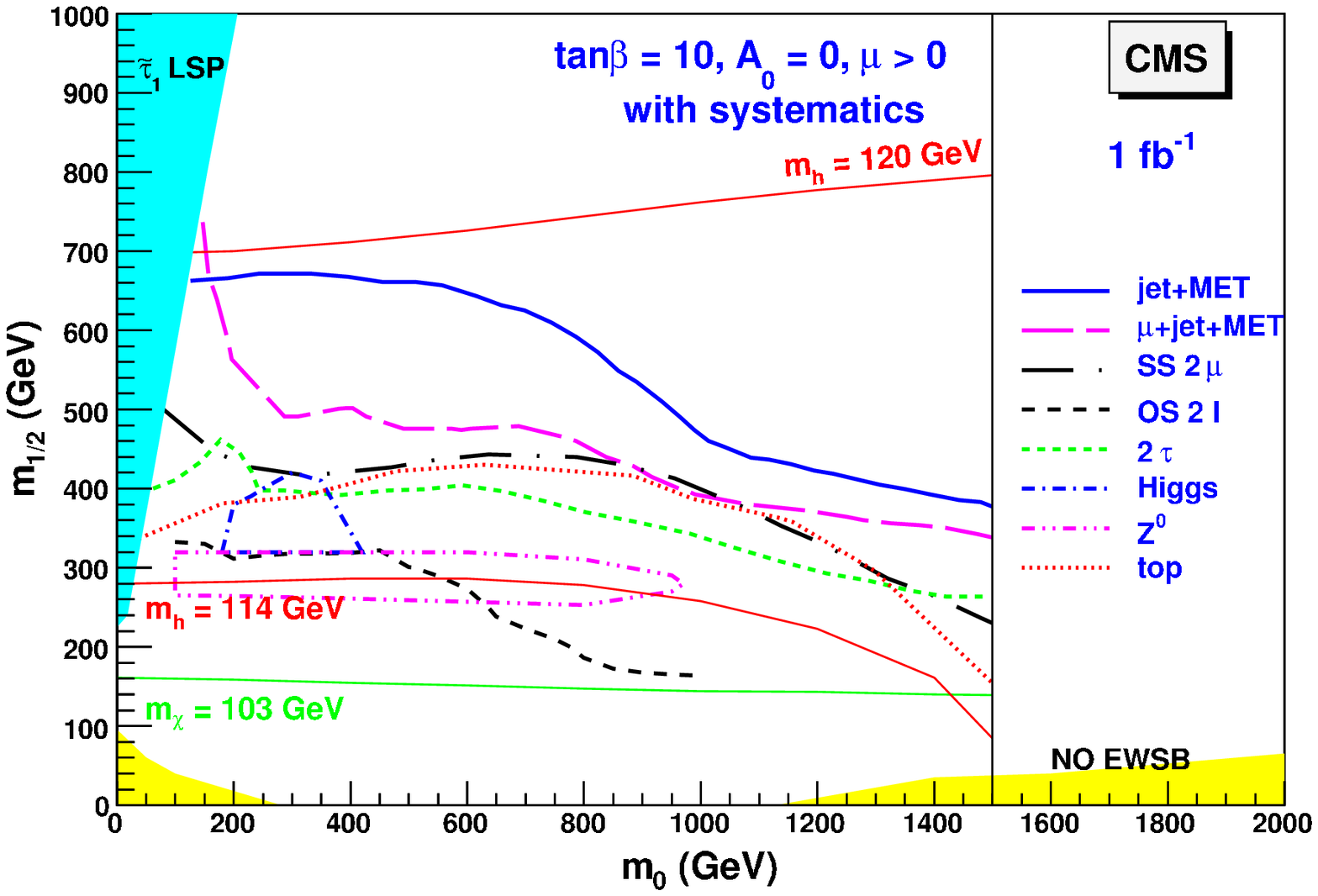} ~~ \\
 A. & B.\\
\vspace{4cm}
\end{tabular}
\vspace{-.4cm}
\caption{The mSUGRA discovery reach in $m_0$-$m_{1/2}$ plane for fixed
$A_0$=0, $tan\beta$=10, $\mu$$>$0 for 1\fb~with systematic uncertainties
denoted by dash lines for the  (A) and with uncertainties already
included for CMS (B). 
\vspace{-.5cm}
}
\label{fig.msugra}
\end{figure}

The ATLAS collaboration studied the mSUGRA model with $n$$\leq$2 leptons
in the final state. The cuts have been optimised with fast S\&R for
{\small $m_0$=100-2000\GeV, $m_{1/2}$=100-1500\GeV, $tan\beta$=5,10,
30,50, $A_0$=0 {\rm and} $\mu$$>$0} model parameters using the 
SUSY-sensitive observables:
{\small
\met, $\rm p_T ^{1^{st} jet}$,
$\rm p_T ^{4^{th} jet}$,
Sphericity$\rm _{\,T}$.} 
The background consists of the following SM processes\,\cite{Mangano:2002ea}:
{\small
$t\bar{t}$+N(0-3)jets,
$W(\ra l\nu)$+ N(2-5)jets,
$Z(\ra l l, \nu\nu)$+N(2-5)jets,
N(2-6) $QCD$ jets.} 
The large cut on \met~{\small (e.g.$>$400\GeV)} effectively removes the SM
background in the wide {\small $m_{1/2}$} region due to mass splitting
between the heaviest and the lightest SUSY particles.  The major
theoretical uncertainties of background cross-sections arise from the low
parton \pt~cut and the small renormalization scale. Experimental
uncertainties of luminosity (5\%), \met~scale (5\%) and jet energy scale
(5\%) are considered. The resulting discovery reach, defined by at least
10 signal events and $S/\sqrt{B}>5$ for 1\fb, is shown in
Fig.\ref{fig.msugra}A. By including uncertainties the discovery potential
curves are lower on {\small $m_{1/2}$} by about 50\GeV~for all channels.
Therefore, ATLAS searches for $n$$\leq$1 lepton channels are sensitive up
to $M_{SUSY}$\,$\sim$\,1.4\TeV~or {\small $m_{1/2}$$\sim$}700\GeV.

The CMS experiment analysed several signatures characteristic for
mSUGRA\,\cite{PTDR2} with the full S\&R and the event selection criteria
optimised for SUSY.  The discovery potential for integrated luminosity of
1\fb~is presented in Fig.\ref{fig.msugra}B. The curves show limits with
all systematic uncertainties included. The inclusive channels,
jet+\met~and $\mu$+jet+\met~yields the best results and allow to probe an
existence of SUSY to the same level of $M_{SUSY}$\,$\sim$\,1.5\TeV~as
ATLAS obtained for $n$$\leq$1 lepton channels.  Other experimental
signatures in the same range of parameters may help to confirm the
discovery. The parameters of mSUGRA can be recovered later from the
measurement of several observables such as the reconstructed masses of
SUSY particle.

\vspace{-.2cm}
\section*{Four-lepton signal from Universal Extra Dimensions}

\begin{figure}[t]
\begin{center}
\begin{tabular}{rr}
\vspace{-5cm}
\epsfxsize=7.0cm \epsfysize=5.0cm \epsfbox{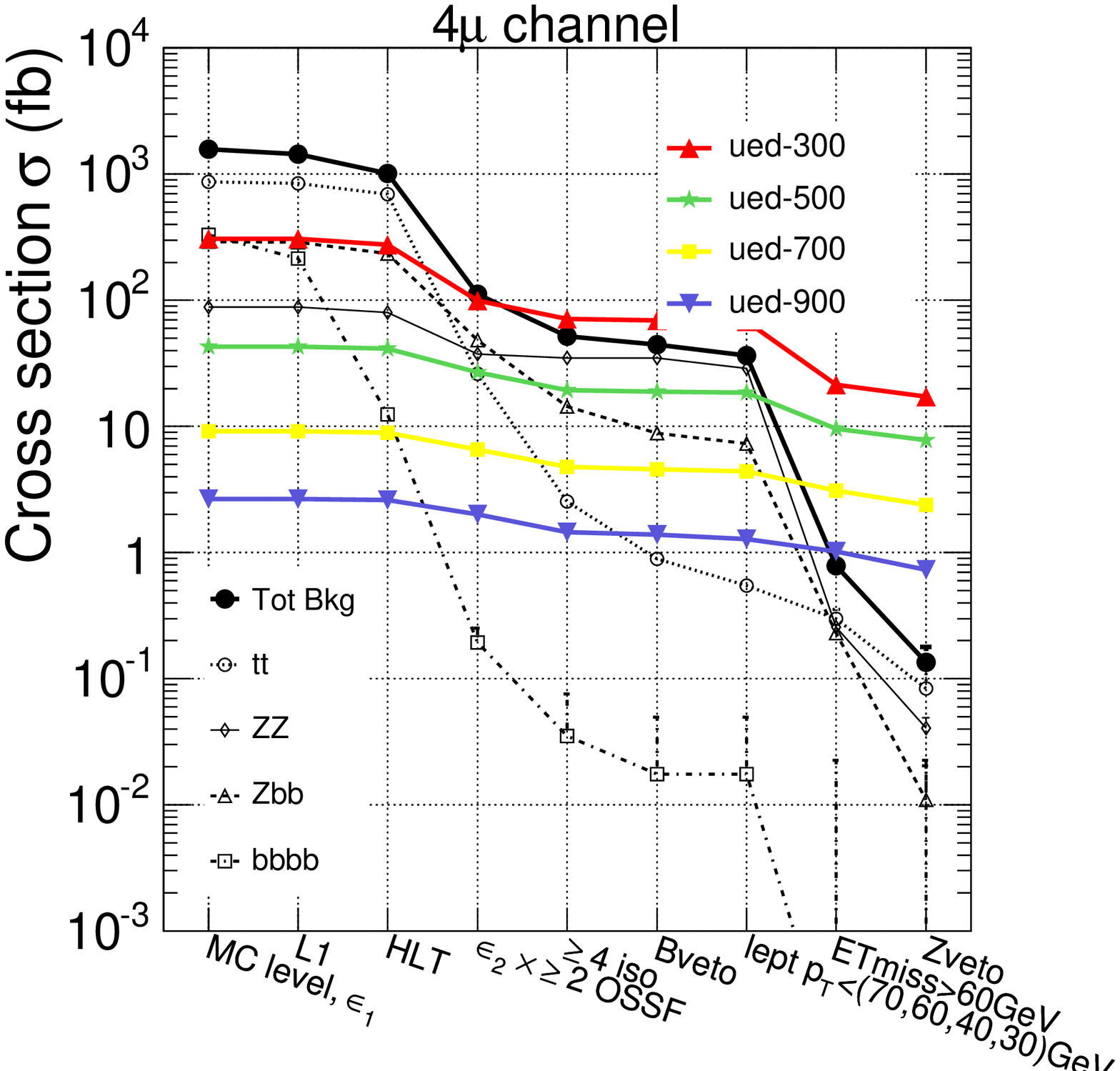}&
                 \epsfysize=5.0cm \epsfbox{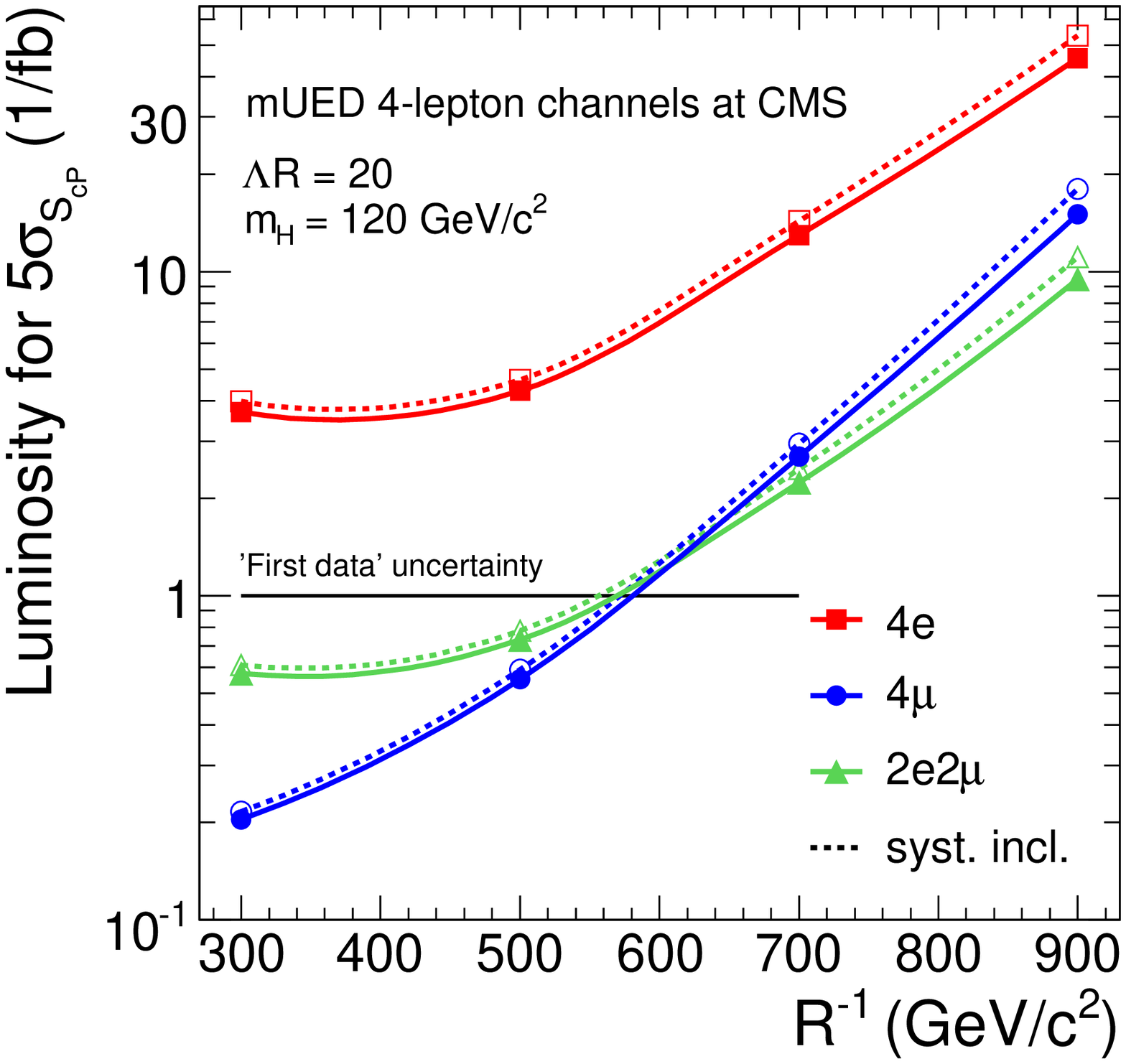}\\
 A. & B. \\
\vspace{4cm}
\end{tabular}
\vspace{-.3cm}
\caption{
The CMS search for mUED signal with the 4-lepton final state.
A. Signal selection efficiency in the 4-muon channel.
B. The mUED 4-lepton discovery potential as a function of the ED size, \Rinv.
\vspace{-.5cm}
}
\label{fig.ued}
\end{center}
\end{figure}

The CMS collaboration studied 4-lepton signatures in the context of
Universal Extra Dimensions\,\cite{Kong:2006pi} (UED) model. The
phenomenology of UED is very similar to that of SUSY, although the origin
of UED comes from the sub-millimetre ED model of the
ADD\,\cite{Arkani-Hamed:1998rs} type. In UED, all SM fields are allowed to
propagate along EDs. Therefore, each SM particle has Kaluza-Klein (KK)
excitations with the same spin contrary to SUSY particles. In the minimal
scenario with only one ED: mUED[\Rinv,\LR], where \Rinv~is a size of the
ED and $\Lambda$ an effective cut-off scale, the KK excitations exhibit
highly degenerate masses even after radiative corrections. The KK-quarks
decay to the lightest and stable KK-photons in a long chain
 ($q^{KK}\ra Z^{KK}q, ~Z^{KK}\ra l^{KK}l^{\pm}, ~l^{KK}\ra
\gamma^{KK}l^{\mp}$)  producing soft leptons and jets in the detector.
Therefore, the 4-lepton final state is considered the best to eliminate
the SM background.  The CMS study\,\cite{Kazana:2006ued} has been
performed with the full S\&R for four sets of mUED parameters (\LR=20 and
\Rinv=300,\,500,\,700,\,900\GeVcc) and for three leptonic channels:
4$\mu$, 4$e$ and 2$e$2$\mu$. Two same-flavour OS isolated lepton pairs
were required in the offline selection in advance of the b-tagging and
Z-veto rejections (Fig.\ref{fig.ued}A). The discovery potential for mUED
in terms of the integrated luminosity needed to achieve signal
significance of 5$\sigma$ (Fig.\ref{fig.ued}B) extends up to
\Rinv=600\GeVcc~for 1\fb data. The systematic uncertainties due to a
limited understanding of the detector performance during initial phase of
the LHC data taking may shift the sensitivity of mUED discovery up to
1\fb.

\vspace{-.2cm}
\section*{Di-lepton resonances from $Z'$ bosons  and Randall-Sundrum Gravitons}
\vspace{-.1cm}

The spin-0 $Z'$~gauge bosons and spin-2 KK excitations of the graviton
with masses of the order of 1\TeV~are predicted by many EDs and GUT's
theories. At LHC, such resonances are produced directly and promptly decay
into pairs of same-flavour OS leptons. Their masses can be measured from
peaks in the invariant mass distribution in the tails of the SM background
processes. This signature has been studied in the CMS experiment.

The dominant background arises from the Drell-Yan (D-Y) lepton pair
production, whereas contributions from $t\bar{t}$ and the vector boson
pair production ($ZZ, ~WZ, ~WW$) are significantly smaller and are highly
suppressed by selection cuts. The K-factors related to the $NNLO$
perturbative $QCD$ calculations are used to correct cross-sections in
function of the di-lepton mass for the D-Y and the new boson production.
Theoretical uncertainties due to a choice of the PDF set and various
experimental uncertainties are also considered including effects of the
misalignment of the muon system in the early ($<1$\fb) and the late
($>$100\fb) phase of the LHC data taking periods. The momentum resolution
of the detector plays a key role in separating the signal from the
background. New reconstruction algorithms have been developed to increase
the lepton reconstruction efficiency. For highly energetic electrons,
their energy deposited as an isolated electromagnetic super-cluster is
corrected for the energy leaking into a hadronic calorimeter and for
electronics saturation effects. For very high-\pt~muons, the track fitting
in the tracker and the muon system are optimised to detect and correct
effects of their energy lost.
\begin{figure}[t]
\begin{center}
\begin{tabular}{rr}
\vspace{-.7cm}
A. & B.~\, \\
\epsfxsize=6.0cm \epsfysize=7.0cm  \epsfbox{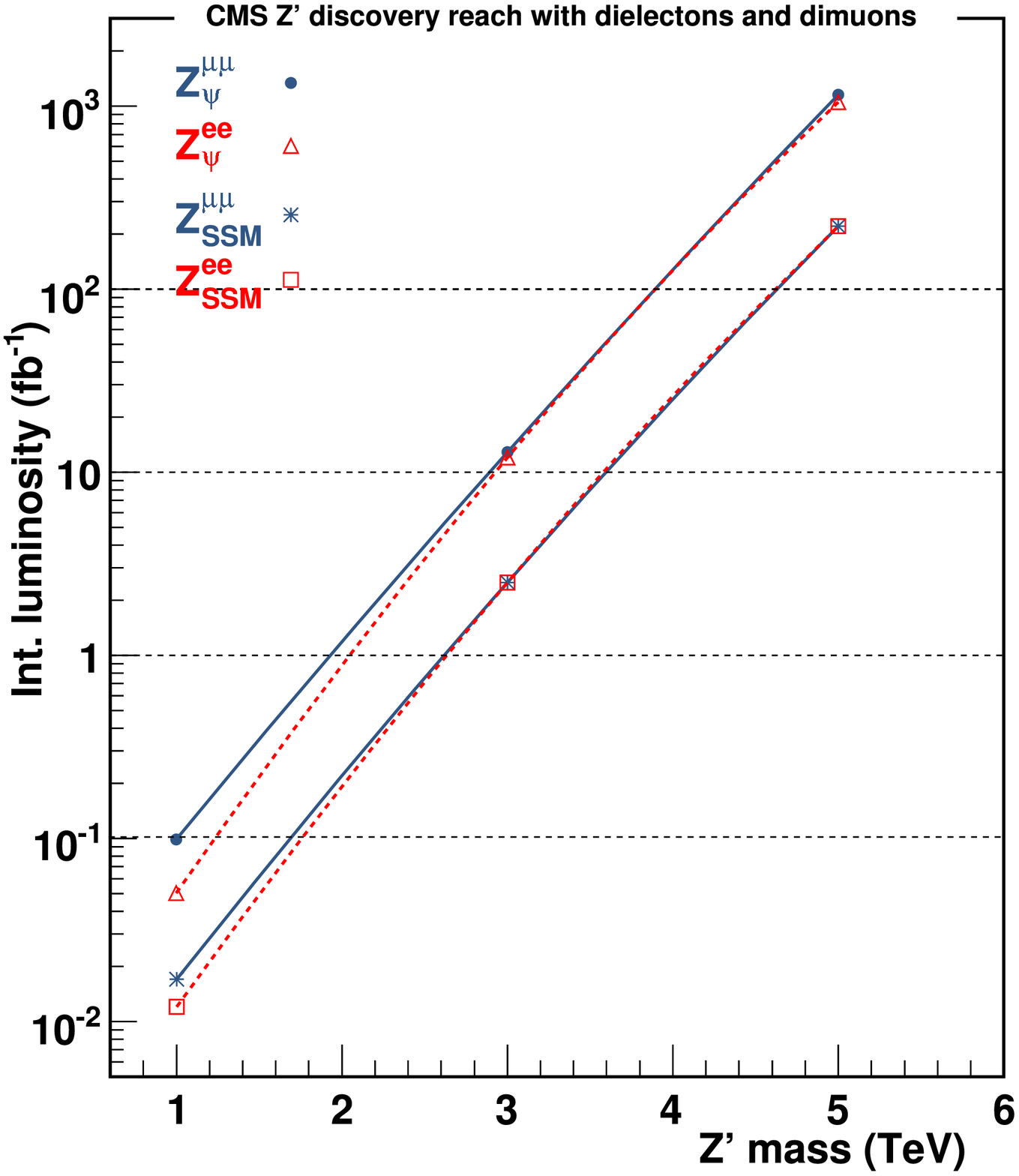} ~~~ &
\begin{tabular}{r}
\epsfxsize=6.0cm \epsfysize=3.3cm  ~\epsfbox{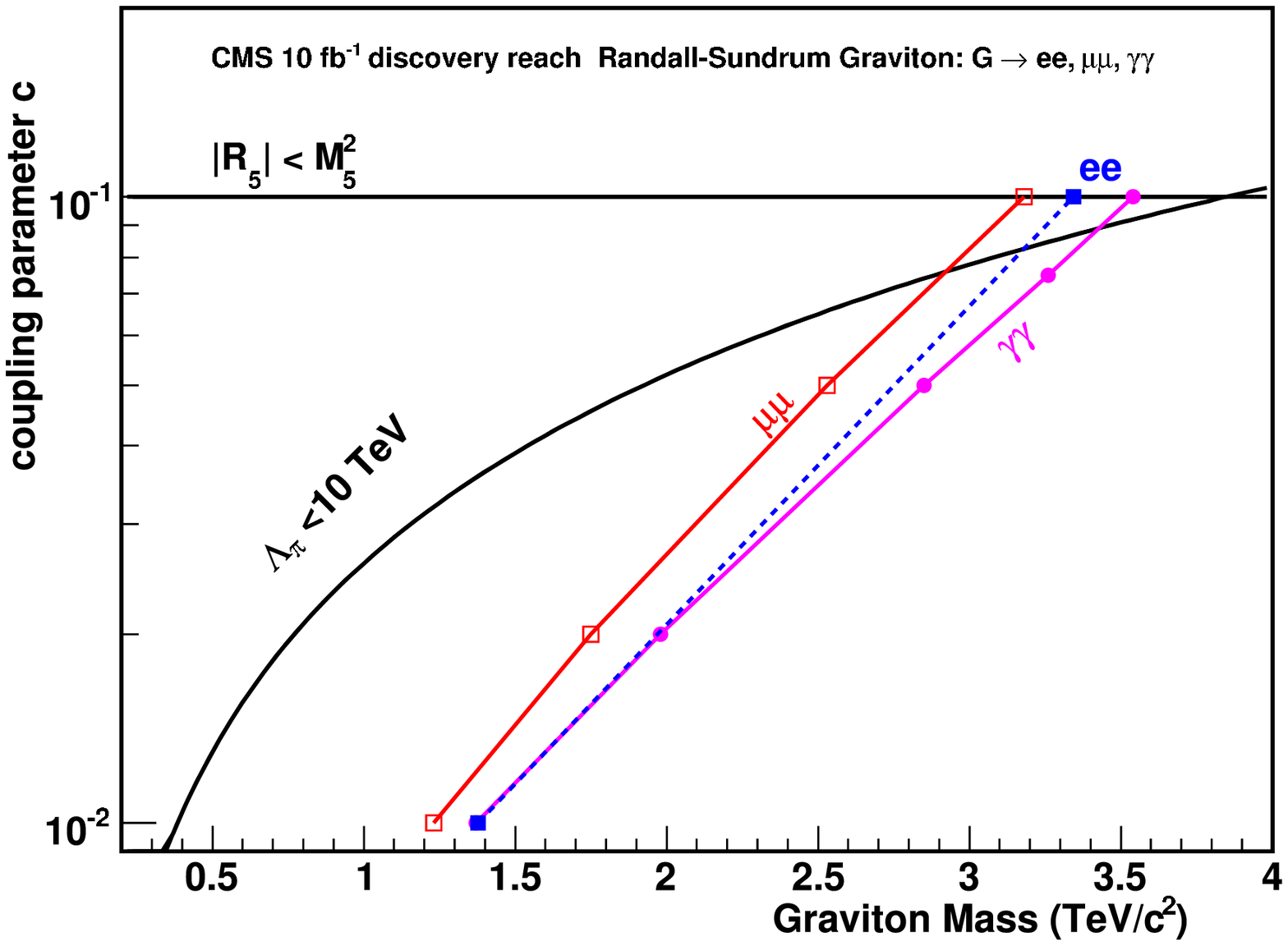}      ~~~~ \\
\vspace{6.3cm}
\epsfxsize=6.0cm \epsfysize=3.3cm  \epsfbox{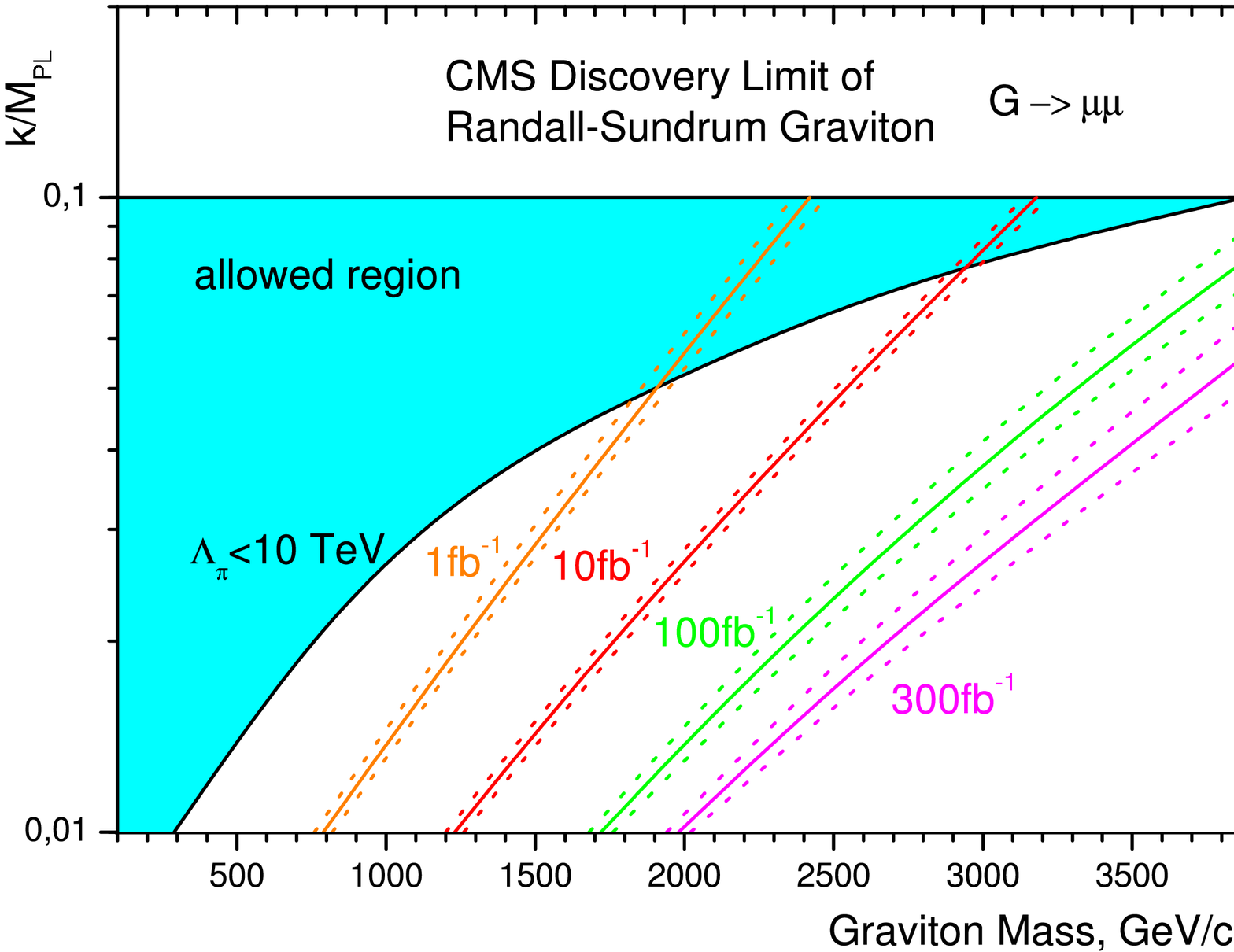}~~~~~ \\
\vspace{-10cm}
~\\
C. \\\vspace{9cm}
\end{tabular}
\end{tabular}
\vspace{-7cm}
\caption{
A. The luminosity required for 5$\sigma$ discovery as a function of $Z'$ mass.
B. The comparison of RS graviton discovery reach for 10\fb~for different
channels.
C. The RS graviton discovery limits with systematic uncertainties (dash lines) for the muon channel.
\vspace{-.5cm}
}
\label{fig.res}
\end{center}
\end{figure}
The results of the CMS analyses\,\cite{PTDR2} obtained with the full S\&R
of signal and background are presented in Fig.\ref{fig.res}.  Both, the
$Z'_{SSM}$ boson originated from the so-called Sequential SM and
$Z'_{\Psi}$ from one of the GUT's models\,\cite{Gunion:1986ky} can be
discovered using 1\fb~of data (above the Tevatron limit of 1\TeV)  up to
$M_{Z'}$\,$\sim$\,$2.6$\TeV~(Fig.\ref{fig.res}A).  Graviton excitations,
$G^*$, are predicted by the Randall-Sundrum\,\cite{Randall:1999ee} (RS)
model in which only one ED is seen by the gravity whereas all SM fields
live in the 3-dimensional Universe.  Couplings and the width of $G^*$ are
given by the parameter $c=k/M_{Planck}$, where $k$\,$\sim$\,$M_{Planck}$
is a mass scale factor.  In Fig.\ref{fig.res}B the RS graviton discovery
reach corresponding to 10\fb~is shown for three decay channels. The muon
channel has the lowest discovery potential due to momentum resolution
effects. Although, the photon channel has the branching ratio almost two
times larger then leptonic ones its reach is comparable to the electron
channel due to $QCD$ and prompt photon irreducible backgrounds. From
Fig.\ref{fig.res}C one can see that 1\fb~of data should be sufficient to
discover RS gravitons decaying to muons with mass up to $M_{G^*}\sim
2.4$\TeV. Experimental methods have been proposed to measure the spin of
such new resonances\,\cite{PTDR2}.

\vspace{-.2cm}
\section*{Summary}

Leptons provide the cleanest signature for exotic searches with the first
LHC data. Thus, inclusive searches with leptons are very promising for
verifying theoretical predictions for physics beyond the Standard Model.
Recent studies on the discovery potential for SUSY performed by the ATLAS
and CMS collaborations are consistent with each other and demonstrate that
probing the production of squarks and gluinos with masses of the order of
1.5\,TeV is possible with integrated luminosity as low as 1\fb. With
1\fb~the searches for EDs can set discovery limits at the level of:  
\Rinv = 600\GeVcc~for mUED model, $M_{G^*}\sim 2.4$\TeV~for the RS
graviton, and $M_{Z'}\sim 2.6$\TeV~for massive $Z'$ bosons.

\end{document}